\documentclass[aps,reprint,twocolumn,nofootinbib,showpacs,floatfix]{revtex4-2}
\usepackage{amsmath,amssymb,bm,graphicx,booktabs}
\usepackage[T1]{fontenc}
\usepackage{placeins,etoolbox}
\AtBeginEnvironment{thebibliography}{\interlinepenalty=10000}
\usepackage[colorlinks=true,linkcolor=blue,citecolor=blue,urlcolor=blue]{hyperref}
\graphicspath{{figures/}}
\usepackage{xcolor}
\AtBeginEnvironment{thebibliography}{\interlinepenalty=10000}

\hypersetup{pdftitle={Amplitude-level NLO predictions for exclusive $J/\psi+\gamma$ production in ultraperipheral collisions},pdfauthor={Author information pending}}

\begin{document}

\title{Amplitude-level NLO predictions for exclusive $J/\psi+\gamma$ production in ultraperipheral collisions}

\author{Xi-Jie Zhan$^{1,2,3}$, Tai-Fu Feng$^{1,2,3}$, Xing-Gang Wu$^4$}

\affiliation{$^1$Department of Physics, Hebei University, Baoding 071002, China}
\affiliation{$^2$Hebei Key Laboratory of High-precision Computation and Application of Quantum Field Theory, Baoding 071002, China}
\affiliation{$^3$Hebei Research Center of the Basic Discipline for Computational Physics, Baoding 071002, China}
\affiliation{$^4$Department of Physics, Chongqing Key Laboratory for Strongly Coupled Physics, Chongqing University, Chongqing 401331, China}

\begin{abstract}

Exclusive $\gamma\gamma\to J/\psi+\gamma$ production is studied in the color-singlet channel at leading order in the heavy-quark velocity. The conventional next-to-leading-order (NLO) result $B+V$, comprising Born production and its interference with the renormalized one-loop amplitude, is compared with $B+V+Q$, where $Q$ retains the complete one-loop square. At photon-photon energies $W=80~\mathrm{GeV}$ and $160~\mathrm{GeV}$, the conventional spectrum becomes negative over finite high-$p_T$ intervals already at renormalization scale $\mu_R=W$, whereas the complete square remains non-negative. In contrast, the total partonic cross section exceeds Born at the larger energies in the energy scan, owing to positive virtual contributions from other scattering directions. For $p$-$p$ collisions at $14~\mathrm{TeV}$ and $Pb$-$Pb$ collisions at $\sqrt{s_{NN}}=5.52~\mathrm{TeV}$, $Q$ raises the central integrated ultraperipheral rates by $11.6\%$--$18.4\%$. Their absolute scale widths over $\mu_R=\{W/2,W,2W\}$ decrease by $36\%$--$67\%$. Retaining the complete one-loop square changes both the predicted integrated rates and the shapes of the differential distributions.

\end{abstract}

\maketitle

Ultraperipheral collisions (UPCs) provide quasireal photon beams while strongly suppressing inelastic hadronic interactions. The equivalent-photon approximation (EPA) connects the electromagnetic fields of relativistic projectiles to photon-induced cross sections~\cite{Budnev:1975poe,Cahn:1990jk}. The high photon luminosity of nuclei and the harder photon spectrum of protons offer complementary opportunities~\cite{Bertulani:2005ru,Baltz:2007kq,Klein:2020fmr}, as demonstrated by the light-by-light scattering measurements~\cite{ATLAS:2017fur,CMS:2018erd}. Heavy quarkonium adds a probe of short-distance QCD and bound-state formation, separated in nonrelativistic QCD (NRQCD) through expansions in the strong running coupling $\alpha_s$ and the relative heavy-quark velocity $v$~\cite{Bodwin:1994jh,Brambilla:2010cs,Lansberg:2019adr}. Quarkonium--photon production has been studied in hadronic collisions~\cite{Drees:1991ig,Doncheski:1993dj,Li:2008ym,Lansberg:2009db} and photoproduction~\cite{Mehen:1996vx,Cacciari:1996zu}.

The exclusive production channel $\gamma\gamma\to J/\psi+\gamma$ is particularly simple: its photons leave no partonic remnants, and at leading order in $v$ the heavy-quark pair is a color singlet. Its NLO QCD correction is known from two-photon studies~\cite{Klasen:2004az,Sun:2017yvt}, recent UPC calculations~\cite{Chen:2025gwp,Jia:2025qul}, and electron--positron collider predictions with additional NRQCD channels~\cite{Jiang:2025miz}. Large negative virtual corrections, including negative spectra in some kinematic and scale choices, are already known in this channel~\cite{Sun:2017yvt,Chen:2025gwp}. Thus a positive integrated rate alone does not establish the reliability of its differential prediction.

The amplitude-level NRQCD prescription developed in the complete NNLO study of $J/\psi\to3\gamma$~\cite{Zeng:2026ois} offers a different organization of the same perturbative information. The physical probability is formed from the coherent amplitude. In inclusive calculations, real and virtual terms generally have to be combined at a common order to cancel infrared singularities. For an exclusive channel whose short-distance amplitude is infrared finite after renormalization and NRQCD matching, one may instead truncate that finite amplitude and retain its complete norm. This keeps the interference among its components and preserves non-negativity. The recent comparison of truncation prescriptions for several exclusive charmonium channels at $B$ factories~\cite{Huang:2026ujq} further motivates differential tests. The crossed process $\gamma\gamma\to J/\psi+\gamma$ probes nontrivial energy and angular dependence, as well as absorptive parts in the physical region. Its complete one-loop norm and the resulting UPC spectra are calculated here.

For $\gamma(k_1)+\gamma(k_2)\to J/\psi(P)+\gamma(k_3)$, the $c\bar{c}$-pair is projected onto ${}^3S_1^{(1)}$ with $p_c=p_{\bar c}=P/2$ and $P^2=M^2=4m_c^2$~\cite{Petrelli:1997ge}. 
Here $m_c$ is the charm-quark mass, and $M=2m_c$ is the $J/\psi$ mass used in the leading-$v$ kinematics.
The hard contributions are
\begin{align}
B&=\tfrac14\sum_{\rm pol}|\mathcal M^{(0)}|^2,\qquad
Q=\tfrac14\sum_{\rm pol}|\mathcal M_{\mathrm{ren}}^{(1)}|^2,\nonumber\\
V&=\tfrac12\Re\sum_{\rm pol}\mathcal M_{\mathrm{ren}}^{(1)}{\mathcal M^{(0)}}^*,\nonumber\\
H_{\mathrm{FO}}&=B+V,\quad H_{\mathrm{AMP}}=B+V+Q,\nonumber\\
H_{\mathrm{AMP}}&=\tfrac14\sum_{\rm pol}|\mathcal M^{(0)}+\mathcal M_{\mathrm{ren}}^{(1)}|^2.
\label{eq:bvq}
\end{align}
$\mathcal M^{(0)}$ is the Born amplitude and $\mathcal M_{\mathrm{ren}}^{(1)}=\mathcal M_{\rm loop}^{(1)}+\mathcal M_{\rm CT}^{(1)}$ is its renormalized one-loop QCD correction, including the counterterm (CT) amplitude. 
The labels FO and AMP denote the conventional fixed-order NLO prediction and the amplitude-level NLO prediction, respectively. 
The long-distance matrix element (LDME), phase space and photon luminosity are excluded from $B,V,Q$.
With $e^2=4\pi\alpha$, the amplitudes scale as $\mathcal M^{(0)}\sim e^3$ and $\mathcal M_{\mathrm{ren}}^{(1)}\sim e^3\alpha_s$. The amplitude-level NLO denotes the full square of $\mathcal M^{(0)}+\mathcal M_{\mathrm{ren}}^{(1)}$: it agrees with the conventional result through order $\alpha_s$ and additionally retains $Q$ at relative order $\alpha_s^2$. This term includes contributions from the absorptive part of the one-loop amplitude and all loop--counterterm interference terms. 
For this exclusive color-singlet channel, the order-$\alpha_s$ correction is purely virtual~\cite{Klasen:2004az,Chen:2025gwp}. The coefficient is finite after on-shell heavy-quark field and mass renormalization and the leading NRQCD projection. Color-octet channels, resolved photons and relativistic corrections are outside the present calculation. 
$Q$ is obtained from the open amplitude:
\begin{equation}
	\mathcal M_{\mathrm{ren}}^{(1)}=\sum_{i=1}^{14}F_iT_i,\qquad
	Q=\tfrac14\sum_{i,j}F_iG_{ij}F_j^*.
	\label{eq:tensor}
\end{equation}
Here $G_{ij}=\sum_{\text{pol}}T_iT_j^*$. There are 14 independent Lorentz structures $T_i$ and renormalized form factors $F_i$, which are kept in a common $D=4-2\epsilon$ representation, where $\epsilon$ is the dimensional regulator, until pole cancellation and the finite four-dimensional limit are taken~\cite{Zeng:2026ois}. 
There are six Born diagrams and 48 nonvanishing projected one-loop contributions for the considered production channel, 
which can be generated with \textsc{FeynArts}~\cite{Hahn:2000kx}. \textsc{CalcLoop}-based routines~\cite{CalcLoop} perform the algebra; integration-by-parts reduction~\cite{Chetyrkin:1981qh, Laporta:2000dsw} with \textsc{Kira}~\cite{Maierhofer:2017gsa, Klappert:2020nbg} yields 31 scalar master integrals. Their physical-region representations are checked against auxiliary mass flow~\cite{Liu:2017jxz, Liu:2022chg}, \textsc{Package-X} and standard one-loop results~\cite{Patel:2016fam, Ellis:2007qk}. 
The one-loop calculation is checked for pole cancellation and incoming-photon exchange symmetry. The result for $Q$ is cross-checked by a direct sum over physical polarization states at a reference phase-space point. 
With the respective reference inputs, the $B,V$ benchmarks reproduce Ref.~\cite{Chen:2025gwp} and the direct color-singlet results of Ref.~\cite{Jiang:2025miz} with $|y_\psi|<2.65$, and agree with the corrected values communicated by the authors of Ref.~\cite{Jia:2025qul}.
The Cauchy--Schwarz inequality gives
\begin{equation}
	V^2\leq4BQ,\qquad
	H_{\mathrm{AMP}}\geq(\sqrt B-\sqrt Q)^2\geq0.
	\label{eq:positivity}
\end{equation}
A common positive integration weight can be included in this inner product, so non-negativity survives phase-space integration and the UPC convolution. In contrast, $H_{\mathrm{FO}}<0$ when $V<-B$. 

In the photon-photon center-of-mass frame, $\theta^*$ is the $J/\psi$ scattering angle relative to the first incoming photon and $c=\cos\theta^*$. The final three-momentum magnitude is $p^*=(W^2-M^2)/(2W)$, so $p_T=p^*\sqrt{1-c^2}$ and $p_{T,\max}=p^*$. $p_T$ denotes the $J/\psi$ momentum transverse to the beam axis.
In the following equations, $X$ stands for one of the three hard contributions $B$, $V$ or $Q$, and $\hat\sigma_X$ denotes its partonic cross-section contribution:
\begin{align}
\frac{\mathrm{d}\hat\sigma_X}{\mathrm{d} c}
&=\frac{W^2-M^2}{32\pi W^4}\langle\mathcal O_1^{J/\psi}({}^3S_1)\rangle\,X,\nonumber\\
\frac{\mathrm{d}\hat\sigma_X}{\mathrm{d} p_T}
&=\frac{p_T}{p^{*2}|c|}
\left[\left.\frac{\mathrm{d}\hat\sigma_X}{\mathrm{d} c}\right|_c+
\left.\frac{\mathrm{d}\hat\sigma_X}{\mathrm{d} c}\right|_{-c}\right].
\label{eq:kinematics}
\end{align}

Both roots $c=\pm\sqrt{1-p_T^2/p^{*2}}$ contribute to the second line. 
The physical predictions are $\hat\sigma_{\mathrm{FO}}=\hat\sigma_B+\hat\sigma_V$ and $\hat\sigma_{\mathrm{AMP}}=\hat\sigma_B+\hat\sigma_V+\hat\sigma_Q$, with the same sums for differential cross sections.
Following the NRQCD factorization, 
the leading-$v$ spin-summed LDME is
\begin{equation}
	\langle\mathcal O_1^{J/\psi}({}^3S_1)\rangle=(2J+1)\frac{2N_c}{4\pi}|R(0)|^2
	=\frac{6N_c}{4\pi}|R(0)|^2,
	\label{eq:ldme}
\end{equation}
where $J=1$, $N_c=3$, and $R(0)$ is the radial wave function at the origin.

At fixed $W$, the plotted differential $K$ factors are
\begin{equation}
K_i(W,p_T)=
\frac{\mathrm{d}\hat\sigma_i(W)/\mathrm{d} p_T}
{\mathrm{d}\hat\sigma_B(W)/\mathrm{d} p_T},
\qquad i=\mathrm{FO},\mathrm{AMP}.
\label{eq:partonicK}
\end{equation}
The horizontal variable $z_T=p_T/p_*$ maps every energy onto $0<z_T<1$; changing from $p_T$ to $z_T$ leaves these ratios unchanged. For the energy scan, $K_i(W)=\hat\sigma_i(W)/\hat\sigma_B(W)$ instead compares the total partonic cross sections.

The UPC cross sections use collinear, unpolarized EPA photons and intact outgoing beam particles. 
Let $\omega_1$ and $\omega_2$ be the energies of the photons emitted by beams 1 and 2 in the collider center-of-mass frame.
Their energies satisfy $W=2\sqrt{\omega_1\omega_2}$ and the pair rapidity is $Y=\tfrac12\ln(\omega_1/\omega_2)$. 
The dimensionless photon luminosity is
\begin{equation}
\begin{aligned}
\mathcal L(\omega_1,\omega_2)
&=\int_{b_1>R_1,\,b_2>R_2}\!\mathrm{d}^2b_1\mathrm{d}^2b_2\\
&\quad\times n_1(\omega_1,b_1)n_2(\omega_2,b_2)P_{\mathrm{no\,had}}(b),
\end{aligned}
\label{eq:luminosity}
\end{equation}
where $\bm b_i$ runs from beam center $i$ to the fusion point, $b=|\bm b_1-\bm b_2|$, and $P_{\mathrm{no\,had}}$ is the probability of no additional inelastic hadronic interaction. Each $n_i$ is a photon number density per $(\mathrm{d}\omega_i/\omega_i)\mathrm{d}^2b_i$.

The electric-dipole form-factor (EDFF) prescription is used, with the beam-radius cuts shown in Eq.~\eqref{eq:luminosity}~\cite{Shao:2022cly,Jia:2025qul}. Its density is
\begin{equation}
n_i(\omega_i,b_i)=\frac{Z_i^2\alpha}{\pi^2}
\frac{\xi_i^2}{b_i^2}
\left[K_1^2(\xi_i)+\frac{K_0^2(\xi_i)}{\gamma_{L,i}^2}\right],
\label{eq:flux}
\end{equation}
where $\xi_i=\omega_i b_i/\gamma_{L,i}$, $Z_i$ is the charge number, $\gamma_{L,i}$ is the beam Lorentz factor, and $K_0,K_1$ are modified Bessel functions. The lengths in Eq.~\eqref{eq:flux} are expressed in natural units.

For $p$-$p$, the elastic-profile survival prescription of Ref.~\cite{Shao:2022cly} is used with slope $b_0$. For $Pb$-$Pb$, $P_{\mathrm{no\,had}}(b)=\exp[-\sigma_{NN}^{\rm inel}A_{\rm nuc}^2
\widehat T_{AA}(b)]$, where $A_{\rm nuc}$ is the nuclear mass number and
$\widehat T_{AA}(b)=\int\mathrm{d}^2r_\perp\,\widehat T_A(r_\perp)
\widehat T_A(|\bm r_\perp-\bm b|)$.
The thickness $\widehat T_A(r_\perp)=\int\mathrm{d} z\,
\rho(\sqrt{r_\perp^2+z^2})$ is obtained from the unit-normalized Woods--Saxon density $\rho(r)=\rho_0/[1+\exp((r-R)/a)]$, with $\rho_0$ fixed by $\int\mathrm{d}^3r\,\rho(r)=1$~\cite{Loizides:2017ack}.

\begin{table}[htb]
\caption{Beam and overlap inputs. $\sqrt{s_{NN}}$ is the collision energy per nucleon pair and $E_{\rm beam}$ is per proton or per nucleon. $R$ is the EDFF radial cutoff and also the $Pb$ matter-density radius, $a$ is its diffuseness, and $b_0$ is the proton elastic slope. $\sigma_{NN}^{\rm inel}$ is the inelastic nucleon--nucleon cross section.}
\label{tab:beams}
\begin{ruledtabular}
\begin{tabular}{lrr}
Input & $p$-$p$ & $Pb$-$Pb$ \\
$\sqrt{s_{NN}}$ [TeV] & 14 & 5.52\\
$E_{\rm beam}$ [GeV] & 7000 & 2760\\
$\gamma_L$ & 7460 & 2962\\
$Z$ & 1 & 82\\
$A_{\rm nuc}$ & 1 & 208\\
$R$ [fm] & 0.7 & 7.1\\
$a$ [fm] & --- & 0.549\\
$\sigma_{NN}^{\rm inel}$ [mb] & 79.04 & 68.50\\
$b_0$ [$\mathrm{GeV}^{-2}$] & 20.6 & ---
\end{tabular}
\end{ruledtabular}
\end{table}

This matter density determines the hadronic overlap, while the photons use the EDFF prescription. The input values are collected in Table~\ref{tab:beams}.

For $X=B,V,Q$, the UPC contribution is
\begin{equation}
\sigma_X=\int\frac{2\mathrm{d} W}{W}\,\mathrm{d} Y\,\mathrm{d} c\,
\mathcal L\!\left(\tfrac W2e^Y,\tfrac W2e^{-Y}\right)
\frac{\mathrm{d}\hat\sigma_X}{\mathrm{d} c}\,\Theta_{\rm sel},
\label{eq:upc}
\end{equation}
where $\Theta_{\rm sel}$ equals one inside the selected phase space and zero outside. The measure follows from $(\mathrm{d}\omega_1/\omega_1)(\mathrm{d}\omega_2/\omega_2)=2\mathrm{d} W\,\mathrm{d} Y/W$. 
The UPC predictions are $\sigma_{\mathrm{FO}}=\sigma_B+\sigma_V$ and $\sigma_{\mathrm{AMP}}=\sigma_B+\sigma_V+\sigma_Q$.

Hard-process parameters: pole mass $m_c=1.5~\mathrm{GeV}$, $M=2m_c$, $\alpha=1/137.065$, and $|R(0)|^2=0.81~\mathrm{GeV}^3$~\cite{Eichten:1995ch}. The two-loop $\overline{\rm MS}$ strong coupling uses $n_f=4$ active flavors with $\Lambda_4=0.326~\mathrm{GeV}$. Selection A imposes $|y_\psi|<4.5$ (no lower $p_T$ cut); selection B adds $p_T>2~\mathrm{GeV}$, where $y_\psi$ is the $J/\psi$ rapidity in the collider frame. Neither selection restricts the final photon rapidity, and fixed-$W$ partonic spectra have no rapidity cuts. 
The central renormalization scale is $\mu_R=W$; total-rate scale variations use $\{W/2,W,2W\}$ and $\{m_T/2,m_T,2m_T\}$ ($m_T=\sqrt{M^2+p_T^2}$), while differential bands employ only the first set. Scales are evaluated at each integration point with other inputs fixed. Numerical errors arise from quadrature convergence, photon-luminosity interpolation and amplitude evaluation; convergence is cross-checked with Gauss--Legendre orders 24 and 36. These numerical uncertainties are separate from renormalization-scale bands. For UPC spectra in $x=p_T$ or $y_\psi$, $K_i(x)=(\mathrm{d}\sigma_i/\mathrm{d} x)/(\mathrm{d}\sigma_B/\mathrm{d} x)$ and $\Delta_Q(x)=(\mathrm{d}\sigma_Q/\mathrm{d} x)/(\mathrm{d}\sigma_{\mathrm{FO}}/\mathrm{d} x)$, with identical cuts in numerator and denominator. For integrated rates, $\Delta_Q=\sigma_Q/\sigma_{\mathrm{FO}}$. The notation $V/B$ or $Q/B$ denotes ratios of integrated or differential contributions for the observable of interest.

\begin{figure*}[htb]
\centering
\begin{minipage}{0.8\linewidth}
\begin{minipage}{0.485\textwidth}\centering\includegraphics[width=\linewidth]{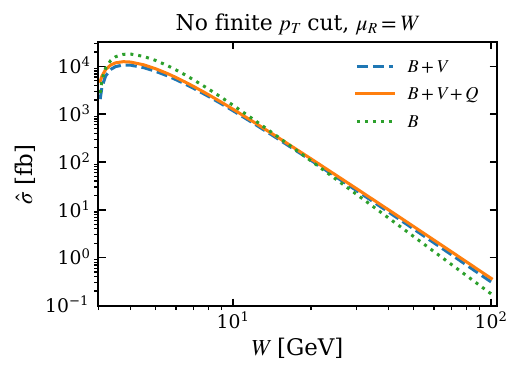}\par{\small (a)}\par\end{minipage}\hfill\begin{minipage}{0.485\textwidth}\centering\includegraphics[width=\linewidth]{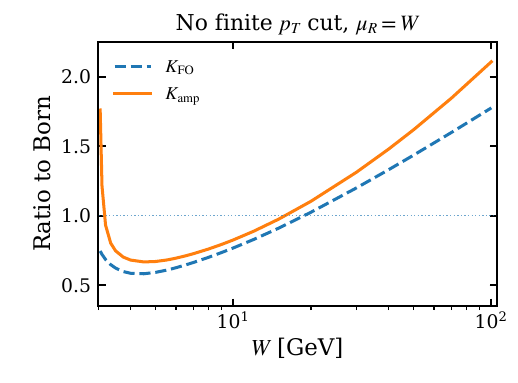}\par{\small (b)}\par\end{minipage}\\[0.6ex]
\begin{minipage}{0.485\textwidth}\centering\includegraphics[width=\linewidth]{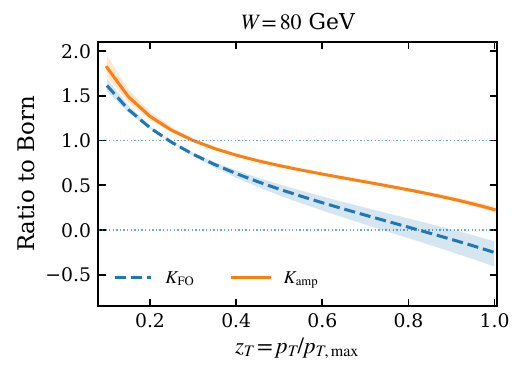}\par{\small (c)}\par\end{minipage}\hfill\begin{minipage}{0.485\textwidth}\centering\includegraphics[width=\linewidth]{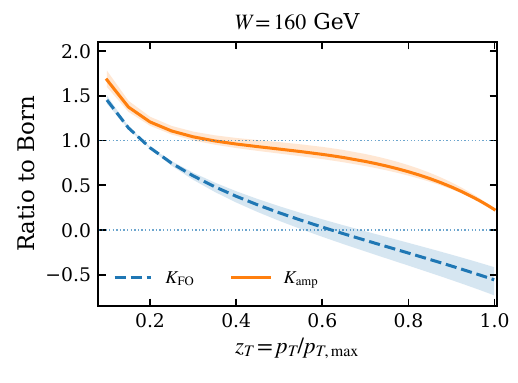}\par{\small (d)}\par\end{minipage}
\end{minipage}
\caption{Partonic production without UPC convolution or rapidity cuts: (a) total cross sections with no finite lower $p_T$ cut and (b) their ratios to Born at $\mu_R=W$; (c),(d) fixed-energy differential $K$ factors versus $z_T=p_T/p_{T,\max}$. In (c),(d), $K_{\mathrm{FO}}$ and $K_{\mathrm{AMP}}$ are the ratios of the corresponding $\mathrm{d}\hat\sigma/\mathrm{d} p_T$ to its Born value at the same $W$ and $p_T$, as defined in Eq.~\eqref{eq:partonicK}. The central curves use $\mu_R=W$ and the bands span $\{W/2,W,2W\}$. The horizontal zero line identifies negative conventional predictions.}
\label{fig:partonic}
\end{figure*}

Figure~\ref{fig:partonic}(a),(b) shows the total partonic cross sections as functions of the specified incoming energy $W$. They rise above threshold and fall at larger $W$. Near threshold, $Q$ can itself be comparable to Born: at $W=3.05~\mathrm{GeV}$, $K_{\mathrm{FO}}=0.746$ and $Q/B=1.015$. At $W=10~\mathrm{GeV}$, the two $K$ factors are instead $0.768$ and $0.826$. The virtual contribution changes sign between the sampled energies $15$ and $20~\mathrm{GeV}$, reaching $K_{\mathrm{FO}}=1.776$ at $W=100~\mathrm{GeV}$; $K_{\mathrm{AMP}}=2.108$ there. With $p_T>2~\mathrm{GeV}$, the threshold becomes $W_{\min}=p_T^{\min}+\sqrt{M^2+(p_T^{\min})^2}=5.606~\mathrm{GeV}$, and the two $K$ factors at $100~\mathrm{GeV}$ are $1.591$ and $1.856$. This cut removes low-$W$ configurations and also restricts the forward/backward angular region at each $W$. These integrated enhancements do not imply a positive virtual correction throughout phase space. The fixed-$W$ spectra in Fig.~\ref{fig:partonic}(c),(d) become negative in finite high-$p_T$ intervals. At $\mu_R=W$, neighboring positive and negative points bracket a sign change at $p_T=32.64$--$32.95~\mathrm{GeV}$ for $W=80~\mathrm{GeV}$, and $49.32$--$50.65~\mathrm{GeV}$ for $W=160~\mathrm{GeV}$. These correspond to $z_T=0.817$--$0.825$ and $0.617$--$0.633$, respectively, so the negative regions are not confined to an arbitrarily narrow endpoint neighborhood. At $W=20~\mathrm{GeV}$ all three scales remain positive on the sampled grid; at $40~\mathrm{GeV}$ only the lower-scale member becomes negative. At $W=80~\mathrm{GeV}$, $p_T=39.9~\mathrm{GeV}$, all three fixed-order $K$ factors are negative: $-0.409,-0.251,-0.125$. Their amplitude-level counterparts are $0.199,0.228,0.262$.

\begin{figure*}[htb]
\centering
\scalebox{0.8}{
\begin{minipage}{0.485\textwidth}\centering\includegraphics[width=\linewidth]{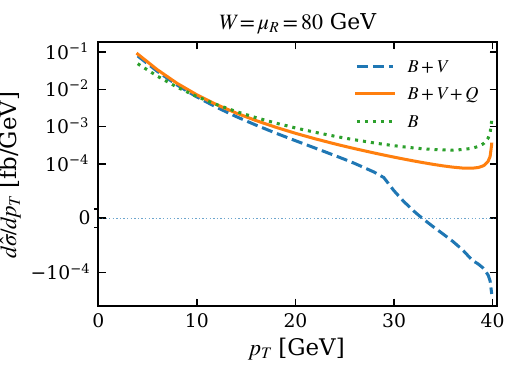}\par{\small (a)}\par\end{minipage}\hfill\begin{minipage}{0.485\textwidth}\centering\includegraphics[width=\linewidth]{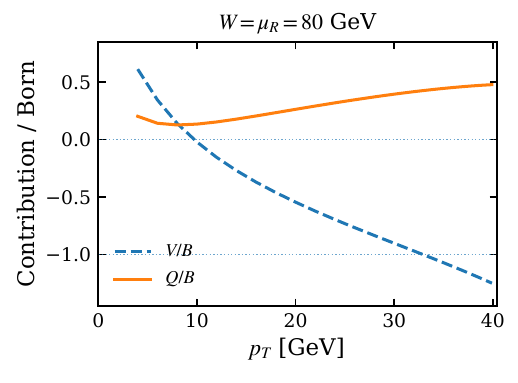}\par{\small (b)}\par\end{minipage}
}
\caption{The $W=80~\mathrm{GeV}$ partonic spectrum at $\mu_R=W$: (a) absolute distributions, using a signed logarithmic scale with a linear interval around zero, and (b) $(\mathrm{d}\hat\sigma_V/\mathrm{d} p_T)/(\mathrm{d}\hat\sigma_B/\mathrm{d} p_T)$ and $(\mathrm{d}\hat\sigma_Q/\mathrm{d} p_T)/(\mathrm{d}\hat\sigma_B/\mathrm{d} p_T)$, labelled $V/B$ and $Q/B$. The rise in the absolute distributions close to $p_{T,\max}$ includes the integrable Jacobian of Eq.~\eqref{eq:kinematics}; it is absent from the ratios.}
\label{fig:W80}
\end{figure*}

Figure~\ref{fig:W80} identifies the source of the negative spectrum. At $p_T=35~\mathrm{GeV}$, $V/B=-1.073$ and $Q/B=0.446$: destructive virtual interference overwhelms the Born contribution, and the full one-loop norm shifts $K_{\mathrm{FO}}=-0.073$ to $K_{\mathrm{AMP}}=0.373$, leaving the differential cross section below the Born level. In the central-scale negative region at $W=160~\mathrm{GeV}$, $Q/B\simeq0.79$--$0.91$, revealing sizable loop corrections. Small $z_T$ yields positive virtual corrections for forward/backward scattering, while large $z_T$ selects central scattering with strongly negative high-energy corrections. Integrating over these competing phase-space regions produces a net integrated enhancement even for negative high-$p_T$ spectra. Raising $W$ at fixed $p_T$ shifts events to forward/backward kinematics, while fixed $z_T$ fixes $|c|$. Though high-energy logarithms are channel dependent~\cite{Bodwin:2014dqa}, these distributions confirm finite-energy negative spectral regions for this exclusive process.

\begin{table*}[htb]
\centering
\caption{Integrated UPC results at $\mu_R=W$. Cross-section columns are in fb for $p$-$p$ and nb for $Pb$-$Pb$. Errors are numerical, not scale errors. $\Delta_Q=\sigma_Q/\sigma_{\mathrm{FO}}$. The last column gives $1-w_{\mathrm{AMP}}/w_{\mathrm{FO}}$, where $w_i=\max_{\mu_R}\sigma_i-\min_{\mu_R}\sigma_i$ within $\mu_R\in\{W/2,W,2W\}$, expressed as a percentage.}
\label{tab:totals}
\begin{ruledtabular}
\begin{tabular}{lcrrrrrrr}
System&Cut&$\sigma_B$&$\sigma_V$&$\sigma_Q$&$\sigma_{\mathrm{FO}}$&$\sigma_{\mathrm{AMP}}$&$\Delta_Q$&Width reduction\\
$p$-$p$&A&155.02&$-58.83$&16.56&$96.19\pm0.19$&$112.75\pm0.23$&17.22\%&63.54\%\\
$p$-$p$&B&14.616&$-5.839$&1.018&$8.777\pm0.026$&$9.795\pm0.029$&11.60\%&36.05\%\\
$Pb$-$Pb$ &A&1009.2&$-386.6$&114.3&$622.6\pm1.3$&$736.9\pm1.5$&18.35\%&66.99\%\\
$Pb$-$Pb$ &B&70.138&$-28.708$&4.918&$41.43\pm0.12$&$46.35\pm0.14$&11.87\%&35.58\%
\end{tabular}
\end{ruledtabular}
\end{table*}

The UPC rates in Table~\ref{tab:totals} show negative integrated interference for both collision systems. The conventional result is $59\%$--$62\%$ of Born; including $Q$ brings it to $66\%$--$73\%$. The additional contribution is $17.2\%$--$18.4\%$ for selection A and $11.6\%$--$11.9\%$ for B. The cut dependence exceeds the beam-species dependence, consistent with removing the low-$W$ region and restricting the allowed angles. Adding $p_T>2~\mathrm{GeV}$ retains $8.69\%$ of the selection-A amplitude-level rate in $p$-$p$ and $6.29\%$ in $Pb$-$Pb$.

The absolute total-rate widths, defined in Table~\ref{tab:totals}, decrease by $35.6\%$--$67.0\%$ in $\{W/2,W,2W\}$ and by $44.6\%$--$70.2\%$ in $\{m_T/2,m_T,2m_T\}$.
For differential spectra, the relative width means the absolute three-scale spread divided by the magnitude of that spectrum at $\mu_R=W$.
The $m_T$ central predictions remain lower than their $W$ counterparts: for example, the amplitude-level $pp$ selection-A rate is 108.86 fb at $\mu_R=m_T$, compared with 112.75 fb at $\mu_R=W$.
At central partonic scattering, $m_T=(W^2+M^2)/(2W)\simeq W/2$ at large $W$; away from it $m_T$ is smaller. The alternative scales can therefore increase the coupling and strengthen the destructive interference, although the net change also depends on $Q$.

\begin{figure*}[htb]
\centering
\begin{minipage}{0.8\linewidth}
\begin{minipage}{0.485\textwidth}\centering\includegraphics[width=\linewidth]{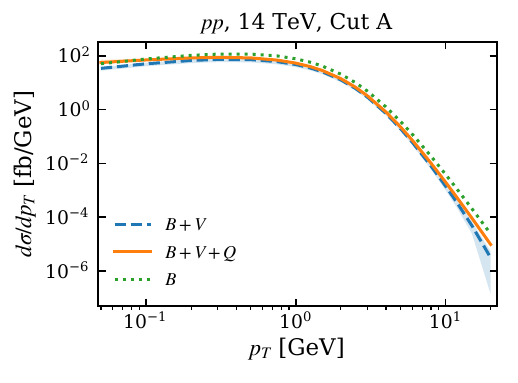}\par{\small (a)}\par\end{minipage}\hfill\begin{minipage}{0.485\textwidth}\centering\includegraphics[width=\linewidth]{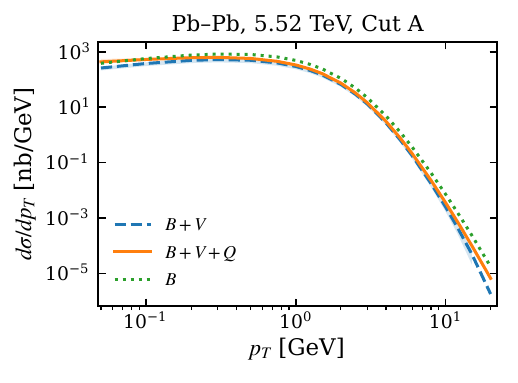}\par{\small (b)}\par\end{minipage}\\[0.6ex]
\begin{minipage}{0.485\textwidth}\centering\includegraphics[width=\linewidth]{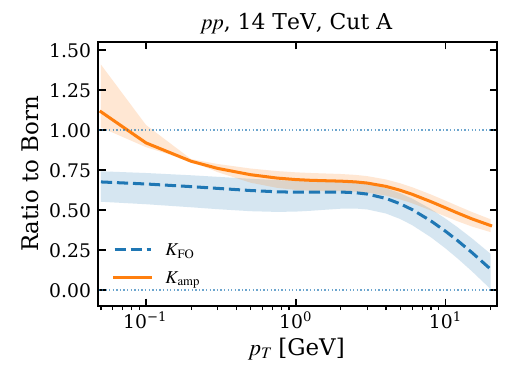}\par{\small (c)}\par\end{minipage}\hfill\begin{minipage}{0.485\textwidth}\centering\includegraphics[width=\linewidth]{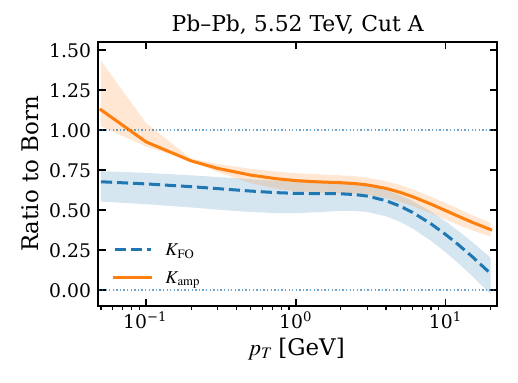}\par{\small (d)}\par\end{minipage}
\end{minipage}
\caption{Selection-A UPC transverse-momentum distributions in $pp$ and Pb--Pb collisions. Upper panels give absolute distributions; lower panels give $K_i(p_T)=(\mathrm{d}\sigma_i/\mathrm{d} p_T)/(\mathrm{d}\sigma_B/\mathrm{d} p_T)$. Central curves use $\mu_R=W$ and bands span $\{W/2,W,2W\}$. Negative scale members cannot appear on the upper positive logarithmic axes; their signs are retained in the lower panels.}
\label{fig:upcpt}
\end{figure*}

The transverse-momentum distributions in Fig.~\ref{fig:upcpt} fall steeply, especially for $Pb$-$Pb$, but the relative effect of $Q$ grows in the tail. At $p_T=1~\mathrm{GeV}$, the fixed-order and amplitude-level values are 46.88 and 52.94 fb/GeV in $p$-$p$, and 296.68 and 335.83 nb/GeV in $Pb$-$Pb$, corresponding to about a $13\%$ increase. At $10~\mathrm{GeV}$ the increases are about $40\%$ and $43\%$. For $Pb$-$Pb$ at $p_T=20~\mathrm{GeV}$, the three fixed-order scale members are $-0.468,1.803,3.550$ fb/GeV; the central numerical error is about 0.007 fb/GeV. The central amplitude-level result is 6.639 fb/GeV. The negative $\mu_R=W/2$ result requires negative partonic contributions, since the convolution weights are non-negative.

At a fixed observed $p_T$, however, the UPC spectrum integrates over the allowed $W,Y$ and both angular branches. A negative fixed-$W$ spectrum alone therefore does not determine the sign of the UPC result. At the lowest displayed point, $p_T=0.05~\mathrm{GeV}$, $\Delta_Q\simeq65\%$--$66\%$ and the relative three-scale width increases from about $28\%$ to $35\%$--$36\%$. Thus narrower total-rate bands are not a general property of the complete square. Very low $p_T$ can also arise from forward scattering at larger $W$, not only from soft-photon kinematics. The present collinear EPA omits photon transverse-momentum smearing and linear-polarization correlations, the latter studied separately in quarkonium UPCs~\cite{Jia:2025oak}.

\begin{figure*}[htb]
\centering
\begin{minipage}{0.8\linewidth}
\begin{minipage}{0.485\textwidth}\centering\includegraphics[width=\linewidth]{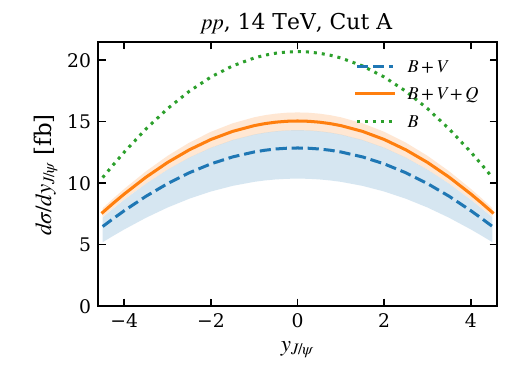}\par{\small (a)}\par\end{minipage}\hfill\begin{minipage}{0.485\textwidth}\centering\includegraphics[width=\linewidth]{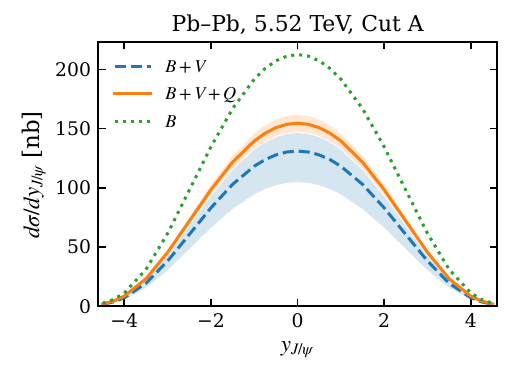}\par{\small (b)}\par\end{minipage}\\[0.6ex]
\begin{minipage}{0.485\textwidth}\centering\includegraphics[width=\linewidth]{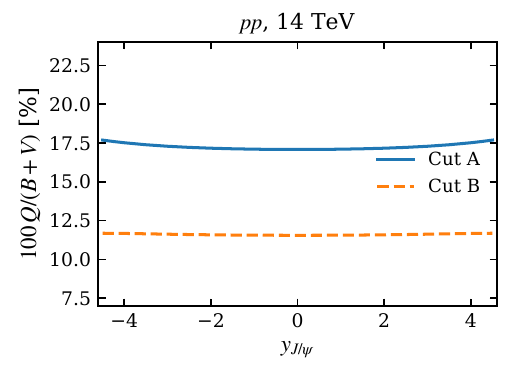}\par{\small (c)}\par\end{minipage}\hfill\begin{minipage}{0.485\textwidth}\centering\includegraphics[width=\linewidth]{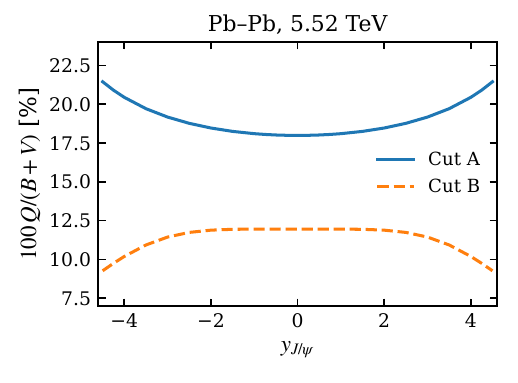}\par{\small (d)}\par\end{minipage}
\end{minipage}
\caption{Selection-A $J/\psi$ rapidity distributions (upper panels), and central relative changes $\Delta_Q(y_\psi)=(\mathrm{d}\sigma_Q/\mathrm{d} y_\psi)/(\mathrm{d}\sigma_{\mathrm{FO}}/\mathrm{d} y_\psi)$ for A and B (lower panels). The upper bands are three-point $W$-scale variations; the lower curves are central ratios, not uncertainty bands.}
\label{fig:rapidity}
\end{figure*}

The rapidity spectra in Fig.~\ref{fig:rapidity} are even for identical beams and symmetric cuts. $Pb$-$Pb$ production is more concentrated near midrapidity. The collider rapidity satisfies $y_\psi=Y+\operatorname{artanh}[p_*c/E_\psi^*]$, with $E_\psi^*=(W^2+M^2)/(2W)$; photon-energy asymmetry and the partonic angle therefore both affect its distribution. In $pp$, the selection-A increase changes only from $17.1\%$ at $y_\psi=0$ to $17.7\%$ near $|y_\psi|=4.5$, and stays near $11.6\%$ for B. Hence the shapes change little. $Pb$-$Pb$ shows more variation: $18.0\%$ to $21.4\%$ for A and $11.9\%$ to $9.3\%$ for B. The relative scale widths decrease throughout the displayed rapidity range for both selections.

\FloatBarrier

In summary, the complete complex one-loop QCD amplitude yields sizable changes in exclusive $J/\psi+\gamma$ production beyond the conventional Born--loop interference. At $W=80$ and $160~\mathrm{GeV}$, the latter makes finite high-$p_T$ intervals negative already at the central scale, although positive contributions at other angles can enhance the total partonic cross section. The full one-loop square preserves non-negativity and raises the central UPC rates by $11.6\%$--$18.4\%$, with nonuniform effects on the differential distributions. It also reduces the total-rate scale widths, but not every differential scale width.

\textit{Acknowledgments.} We thank the authors of Ref.~\cite{Jia:2025qul} for kindly sharing their computational code for comparison with our results. This work was supported partly by the National Natural Science Foundation of China under Grant Nos.12305083, 12575080, and 12547101.

\bibliographystyle{iopart-num}

\bibliography{references}

\end{document}